\documentstyle[12pt,epsfig]{article}

\newcommand{\be}{\begin{equation}}
\newcommand{\ee}{\end{equation}}

 1
\font\elevenrm=cmr10  scaled\magstep 1
 1

\def\NOE{{\em N\raise.5ex\hbox{O \kern-0.47em}E\kern.4em}}

\begin{document}

\hspace{7cm}
{\bf Neutrino Oscillation Workshop (NOW'98)}

\hspace{8cm}
{\bf September 7-9, 1998, Amsterdam}

\vspace*{2cm}
  \centerline{\bf {\LARGE NOE, a Neutrino Oscillation Experiment}}
  \centerline{\bf {\LARGE at Gran Sasso Laboratory}}
\vspace{1cm}
\centerline{Paolo Bernardini $^*$, Giancarlo Barbarino $^\dag$ and 
Fausto Guarino $^\dag$}
\vspace{0.3cm}
\centerline{(\NOE Collaboration)}
\vspace{0.8cm}
  \centerline{* Dipartimento di Fisica dell'Universit\`a and INFN}
  \centerline{\elevenrm via per Arnesano, 73100 Lecce, Italy}
\vspace{0.3cm}
  \centerline{\dag \ Dipartimento di Scienze Fisiche dell'Universit\`a and INFN}
  \centerline{\elevenrm via Cintia, 80126 Napoli, Italy}
\vspace{2cm}

\begin{abstract}
The project of a large underground experiment (\NOE) devoted to long 
baseline neutrino oscillation measurement is presented. The apparatus 
is composed by calorimetric modules interleaved with TRD modules and 
has been optimized to be sensitive in the region of $\sin^2 2\theta$ 
and $\Delta m^2$ suggested by the atmospheric neutrino oscillation 
signal.
\end{abstract}
\vspace{0.7cm}

\section{Introduction}
The first version of the \NOE proposal has been submitted to the Gran 
Sasso Scientific Committee in February 1996 \cite{noe0}. From that 
time on, many efforts (technical design, tests, simulations and so on) 
have been done to improve the project, in terms of scientific goals,
detector performances and granularity, safety and reduction in costs 
\cite{noe1, noe2, noe4}. Here a brief progress report is presented, 
mainly devoted to describe the latest developments with respect to 
\cite{noe6}.

The scientific goal of \NOE long baseline (LBL) experiment is the 
measurement of neutrino masses looking for $\nu_\mu \rightarrow \nu_e$ 
and $\nu_\mu \rightarrow \nu_\tau$ oscillations. The strategy of \NOE 
design is to have oscillation sensitivity by looking for the $\tau$ 
decay ($\nu_\mu \rightarrow \nu_\tau$ oscillation) or for an electron 
excess ($\nu_\mu \rightarrow \nu_e$ oscillation) and by measuring a 
deficit of muons in apparent NC/CC ratio.

The major experimental hint for the $\nu$ oscillation search in the 
region of low $\Delta m^2$ ($10^{-3} \div 10^{-2} \ eV^2$) comes from 
the muon deficit observed in atmospheric neutrino flux measurements 
\cite{soudan, macro1, superk}. Recent results from LBL reactor 
neutrino experiment (CHOOZ) exclude neutrino oscillation in $\bar\nu_e$ 
disappearance mode in the range $\sin^2 2\theta > 0.18$ for large 
$\Delta m^2$ \cite{chooz}.

Given the confirmations of the atmospheric neutrino anomaly and the 
negative CHOOZ result, a LBL experiment has to fulfill the following 
requirements:

\begin{enumerate}
\item {\bf $\nu_\tau$ tagging.} 
The search for $\nu_\tau$ appearance becomes fundamental
in order to confirm the oscillation phenomenon. This search requires 
detector high performances to reveal the $\tau$ decay in different
channels.

\item {\bf Measurement of the ratio NC/CC.} This robust and 
unambiguos test is mandatory to investigate on the existence of a 
neutrino oscillation signal. Moreover in a scenario, never ruled out, 
in which a neutrino oscillates in a sterile neutrino the oscillation 
can be discovered only by measuring a deviation from expected NC/CC 
ratio. There is no doubt that this measurement can be done only with 
a massive detector.

\item {\bf  Atmospheric neutrinos.} After the last results which 
suggest smaller values of $\Delta m^2$, the interest for the 
atmospheric neutrinos is raised up. It would be interesting to test
this effect using a massive apparatus based on a different technique 
with respect to the water \v{C}erenkov detectors.

\item {\bf Fast response.} If a beam from CERN to Gran Sasso will be
available in the next years, a strong competition with American and 
Japanese LBL programs is foreseen: at present, the $7 \ kton$ \NOE project 
can adequately compete with the $8 \ kton$ MINOS detector and with K2K.
\end{enumerate}

According to these remarks the \NOE program can be summarized 
in this way:

\begin{itemize}
\item Direct $\nu_\tau$ appearance by kinematical $\tau$ decay 
reconstruction and inclusive (NC/CC) $\nu_\mu$ disappearance.

\item Investigation of $\nu_\mu \rightarrow \nu_e$ oscillation in
a mixing angle region two orders of magnitude beyond the CHOOZ limit.

\item Atmospheric neutrino studies.
\end{itemize}

In order to improve the $\nu_\tau$ search, the apparatus has been 
implemented with Transition Radiation Detector (TRD) interleaved 
between calorimetric modules (CAL). The combination of TRD and CAL 
informations strongly enforces $e$, $\mu$, and $\pi$ identification, 
thus permitting the study of the $\tau$ decay. In particular the 
$\tau \rightarrow e\nu\nu$ channel is detectable with a clean 
signature because of the low background (residual $\nu_e$ beam). 
Moreover the good electron identification in the TRD and the low 
$\pi^o$ background allow to reach high $\sin^2 2\theta$ sensitivity 
looking for $\nu_\mu \rightarrow \nu_e$ oscillation, thus considerably 
enlarging the region investigated by CHOOZ.

\begin{figure}[t]
\hspace{2cm}
\epsfig{figure=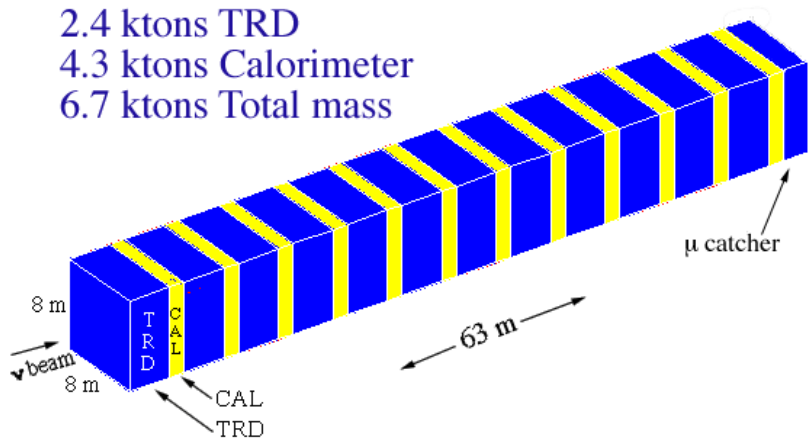,width=10cm,bbllx=31bp,bblly=50bp,bburx=550bp,bbury=300bp}
\vspace{1cm}
\caption{Schematic view of the \NOE detector \label{fig:noex}} 
\end{figure}

\section{The \NOE detector}
\begin{figure}
\hspace{3cm}
\epsfig{figure=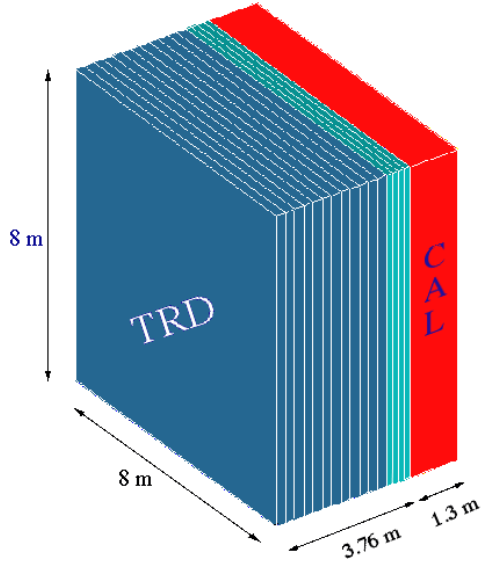,width=8cm,bbllx=31bp,bblly=50bp,bburx=550bp,bbury=550bp}
\vspace{-1cm}
\caption{The Basic Module of the \NOE detector \label{fig:detm}} 
\end{figure}

The detector (Fig.~\ref{fig:noex}) is a classical fixed target 
apparatus consisting of a sequence of 12 modules. Each one is composed 
by a lighter part (TRD target) in which vertex, $e$-identification and
kinematics are defined, followed by a fiber scintillating calorimeter 
devoted to absorb the event and to measure the energy. The Basic 
Module (BM) is shown in Fig.~\ref{fig:detm}. Appearance measurement
is performed on the events generated in TRD ($2.4 \ kton$),
disappearance measurement on all events (generated in $\sim 7 \ kton$
of target).

The calorimetric element is an $8 \ m$ long bar having iron ore and 
scintillating fibers\footnote{Extruded scintillator strips with 
wavelength shifter fibers have been also studied.} suitably 
distributed inside and parallel to the axis. The calorimetric module 
is made by alternate planes of crossed bars. The calorimetric bar 
consists of more logical cells with square cross section, the 
iron/iron-ore volume ratio is 0.3. The iron ore is radiopure and 
practically cost free.

At present the latest production developments of $2 \ mm$ diameter
scintillating fibers provides an attenuation length of $4.5 \ m$
and an increase in light yield of $10 \div 15 \ \%$. These figures 
allow to build $8 \ m$ long bars. Further investigations to improve 
fiber features are in progress because longer fibers shall permit to 
enlarge the \NOE cross section ($9 \times 9 \ m^2$) and mass 
($8 \ kton$). It is worth noting the very high intrinsic granularity 
of the proposed calorimeter : the average distance between the fibers 
inside the absorber is of the order of $3 \ mm$. In a simple and 
often used way, all fibers are grouped together at each side of the 
calorimetric bar and sent to single or multipixel photodetector.

The TRD module consists of 32 vertical layers of $8 \times 8 \ m^2$ 
area, each made by polyethylene foam radiator ($\rho \sim 100 \ mg/cm^3$) 
and 256 proportional tubes ($3 \times 3 \ cm^2$ cross section), filled 
with an $Ar$ (60\%) - $Xe$ (30\%) - $CO_2$ (10\%) mixture, already tested 
in MACRO experiment. Consecutive layers have tubes rotated of $90^\circ$.

A graphite wall of $5 \ cm$ thickness is set in front of each of the
first 24 layers of the TRD module acting as a $174 \ ton$ target for 
$\nu_e$ and $\nu_\tau$ interactions, to be identified in the following 
layers. The last target wall is followed by 8 TRD layers in order 
to identify the secondary particles. Each target wall corresponds to 
$0.25 \ X_0$ while the entire TRD basic module corresponds to about 
$7 \ X_0$ and $3.5 \ \lambda_I$. The total length is about $3.76 \ m$.

So many layers of proportional tubes permit to determine the muon 
energy by means of multiple measurements of energy loss $dE/dx$.
Combining informations coming from both subdetectors (TRD and CAL) the 
discrimination between $e$, $\mu$ and $\pi$ is largely enforced allowing 
the study of several channels for the neutrino oscillation signature.

\section{$\nu_\tau$ appearance and requirements about $\nu$ beam}
The rate of $\nu_\tau$ CC events is given by
\be R_\tau = A \int \sigma_\tau P_{osc} \Phi dE, \ee
where E is the energy, $\sigma_\tau$ the $\nu_\tau$ CC cross section, 
$P_{osc}$ the oscillation probability, $\Phi$ the muon neutrino flux
and $A$ the number of target nucleons in the detector. The search for 
$\nu_\tau$ requires that the term $\sigma_\tau P_{osc} \Phi$ is large. 
Therefore a dedicated $\nu$-beam has to provide most of its flux in 
the energy range where the factor $\sigma_\tau P_{osc}$ is larger.

\begin{figure}
\begin{tabular}{cc}
\epsfig{file=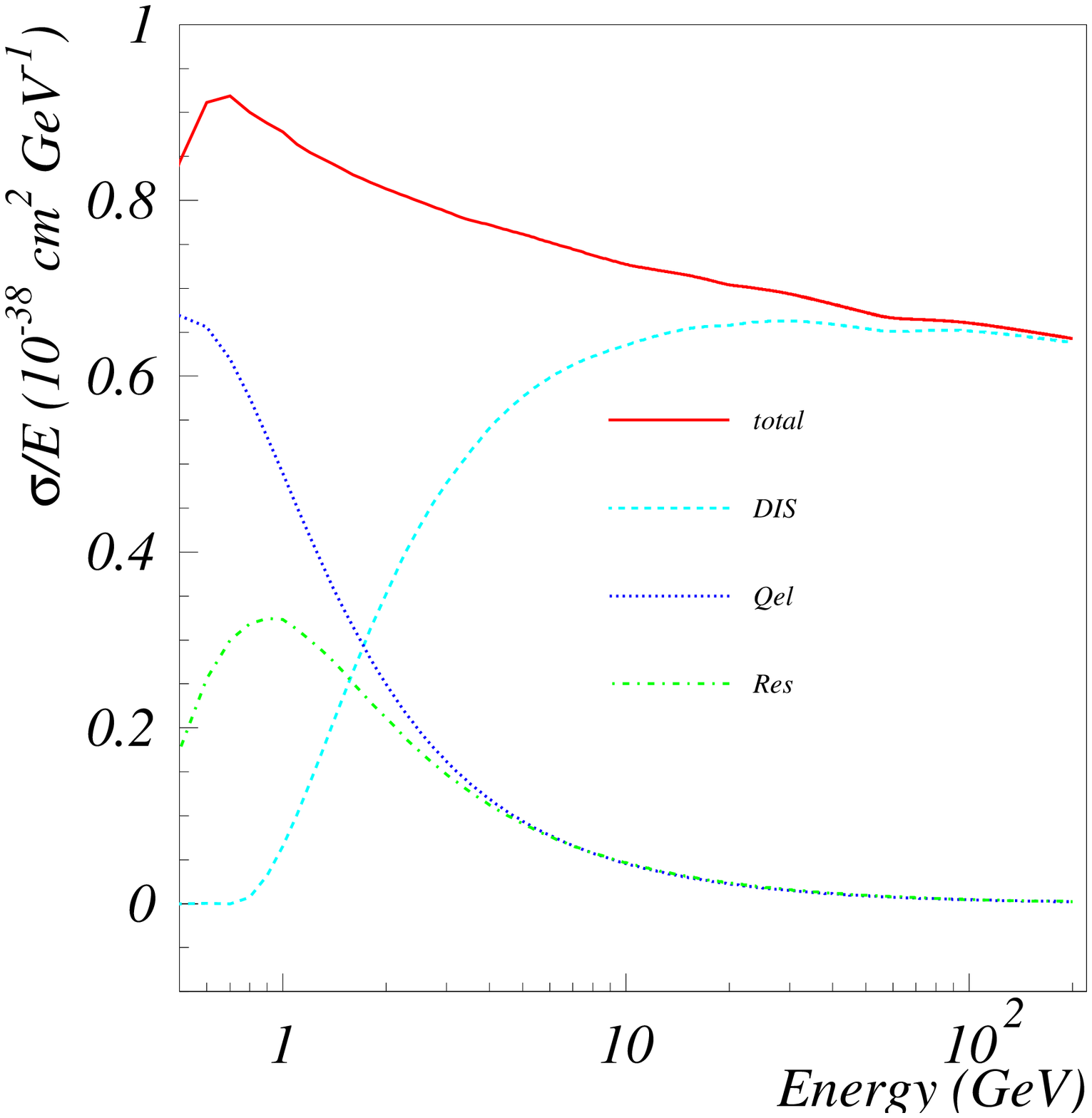,width=7.5cm,bbllx=31bp,bblly=150bp,bburx=550bp,bbury=630bp}
&
\epsfig{file=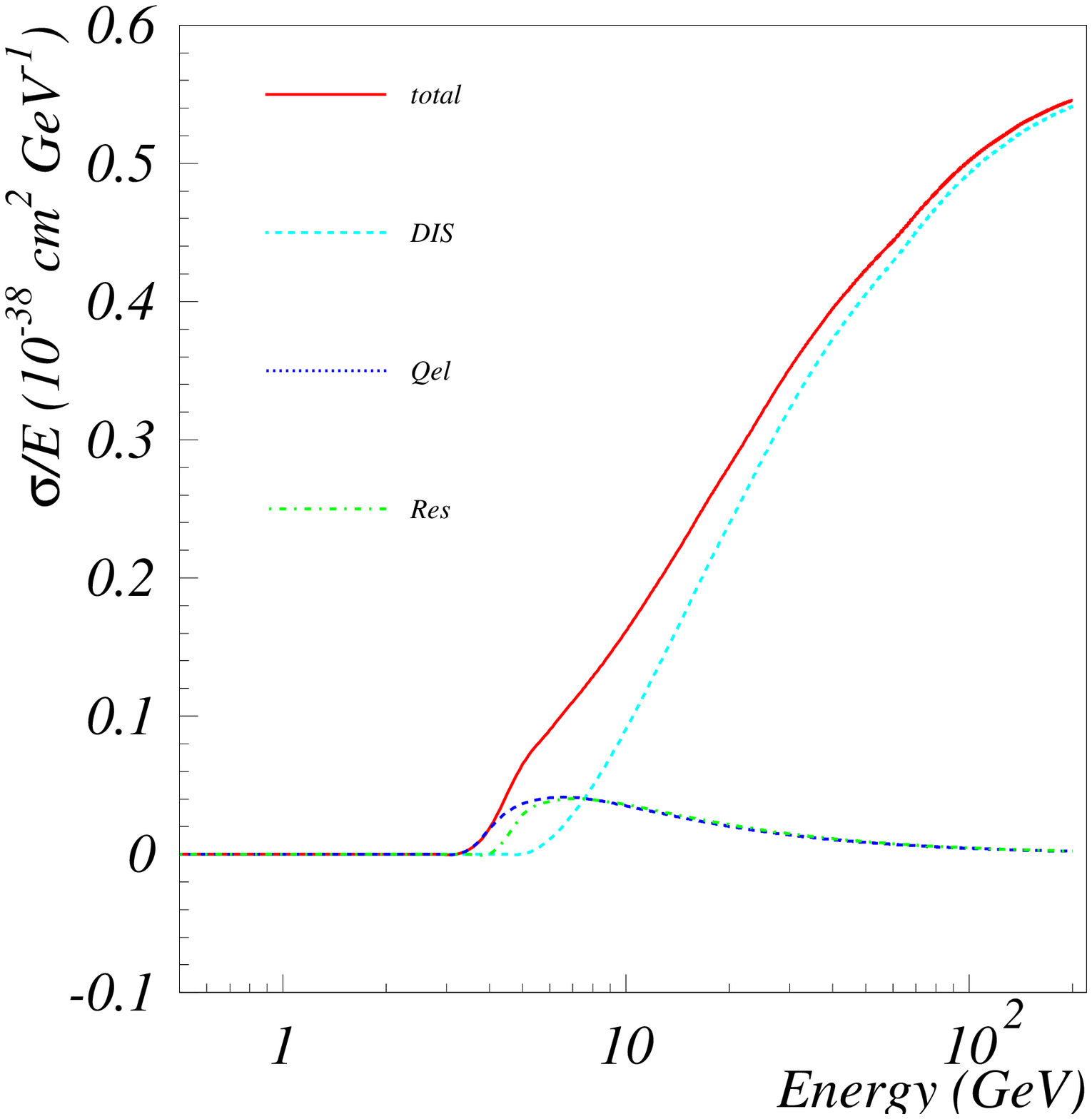,width=7.5cm,bbllx=31bp,bblly=150bp,bburx=550bp,bbury=630bp}
\end{tabular}
\caption{\label{fig:sigma} Ratio of $\nu_\mu$ and $\nu_\tau$ CC cross 
	section over energy.}
\end{figure}

\begin{figure}
\vspace{2.5cm}
\hspace{3.5cm}
\epsfig{file=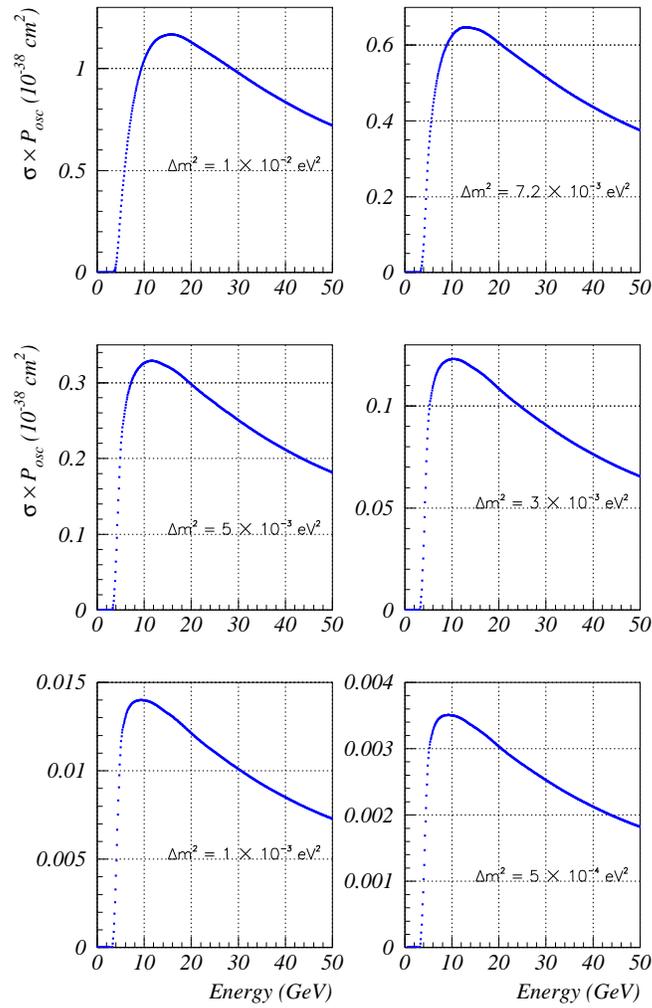,width=9cm,bbllx=60bp,bblly=40bp,bburx=550bp,bbury=630bp}
\caption{\label{fig:p_per_sigtau} $\nu_\tau$ CC cross section multiplied 
 by the oscillation probability for different values of $\Delta m^2$.}
\end{figure}

Assuming the mixing of two neutrinos, the oscillation probability results
\be P_{osc} = \sin^2 2\theta \sin^2 (1.27 \Delta m^2 L/E), \ee
where $L = 731 \ km$ is the distance CERN -- Gran Sasso. CC cross section 
as a function of energy is shown in Fig.~\ref{fig:sigma} for $\nu_\mu$ 
and $\nu_\tau$. We have to note that the $\tau$ cross section grows 
slowly with energy above a threshold of about $3.5 \ GeV$. The factor 
$\sigma_\tau P_{osc}$ is shown in Fig.~\ref{fig:p_per_sigtau} for 
different values of $\Delta m^2$. The optimal energy is about $15 \ GeV$ 
for $\Delta m^2 = 0.01 \ eV^2$ and decreases gently with $\Delta m^2$ 
towards a limiting value of about $10 \ GeV$.

\begin{figure}
\hspace{3.5cm}
\psfig{file=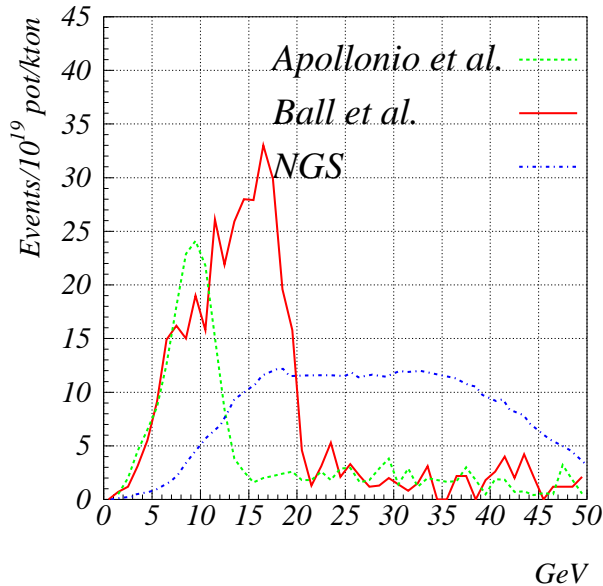,width=8cm,bbllx=31bp,bblly=180bp,bburx=550bp,bbury=630bp}
\caption{\label{fig:spettri} $\nu_\mu$ CC interaction spectra
        for different proposed beams.}
\end{figure}

\begin{figure}[ht]
\begin{tabular}{cc}
\epsfig{file=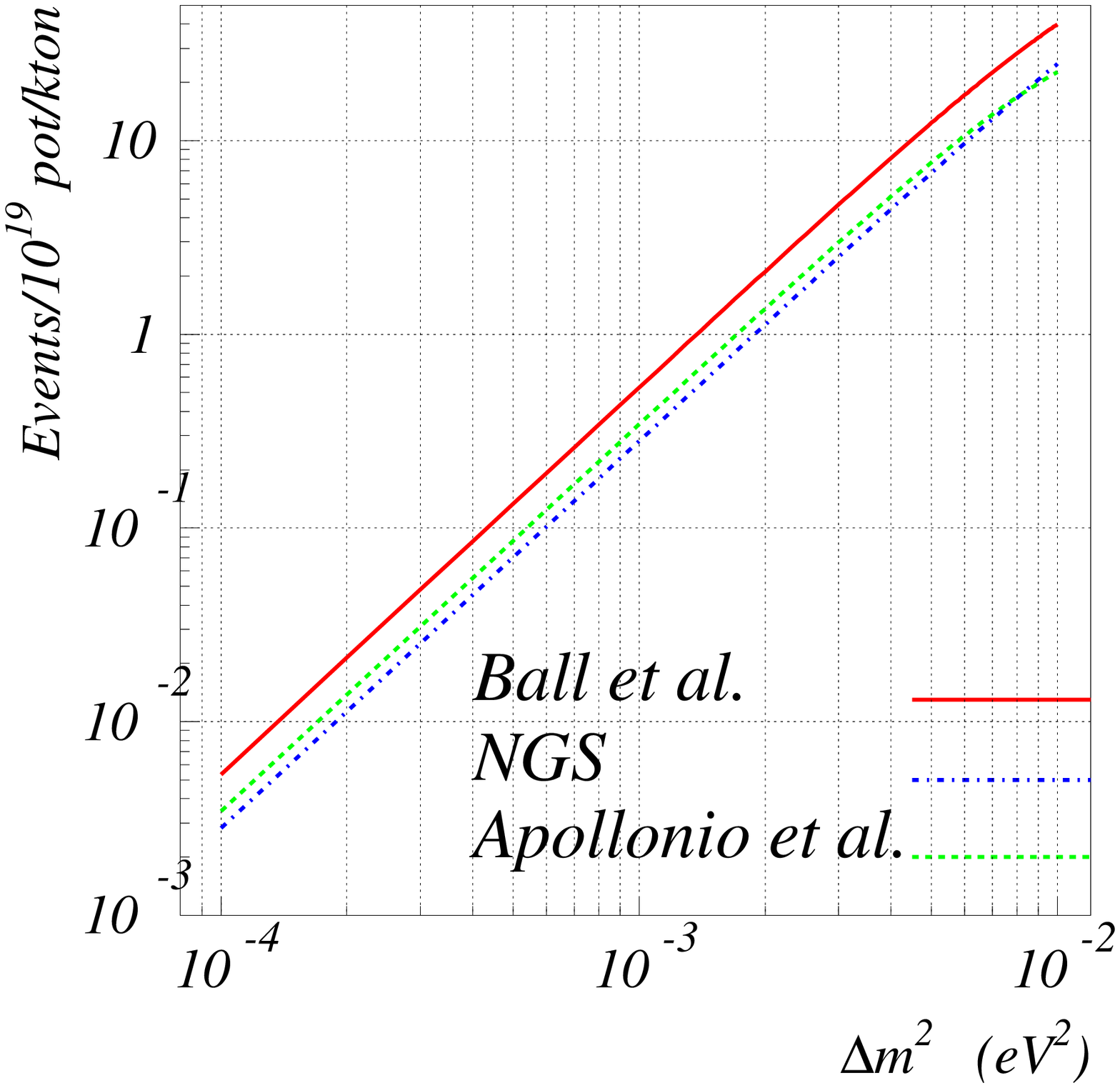,width=8.2cm,bbllx=31bp,bblly=180bp,bburx=550bp,bbury=630bp}
&
\epsfig{file=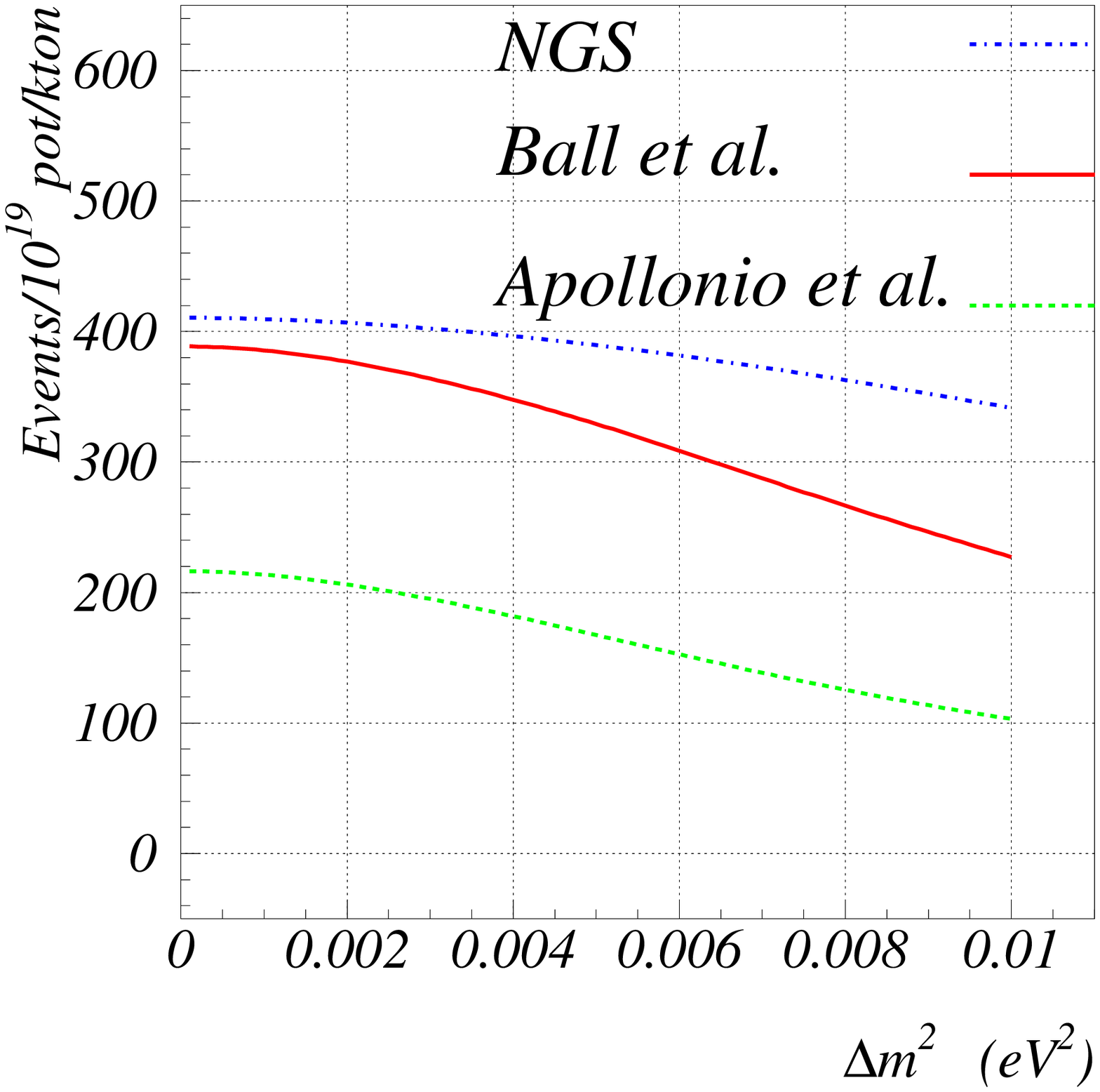,width=8.2cm,bbllx=31bp,bblly=180bp,bburx=550bp,bbury=630bp}
\end{tabular}
\caption{\label{fig:prod} First plot - $\nu_\tau$ CC interactions for
  different beams vs $\Delta m^2$. Second plot - residual $\nu_\mu$ 
  CC interactions for different beams vs $\Delta m^2$.}
\end{figure}

As an example the number of $\nu_\tau$ CC interactions per $10^{19}$ 
protons on target per $kton$ has been calculated for different proposed 
beams \cite{ball, ngs, bald} whose spectra are shown in 
Fig.~\ref{fig:spettri}. The rate of $\nu_\tau$ CC interactions is shown 
in the first plot of Fig.~\ref{fig:prod} as a function of $\Delta m^2$. 
The highest $\tau$ yield is obtained with a beam like that proposed by 
Ball {\em et al.} \cite{ball}, narrowly peaked around an energy that 
better matches the appearance requirements.

Fig.~\ref{fig:prod} shows also the rate of residual $\nu_\mu$ CC 
interactions. From comparison of the plots in Fig.~\ref{fig:prod} we 
observe that the beam of Ref.~\cite{ball} has a higher ratio of 
oscillated to unoscillated events (the best signal to noise ratio) 
when compared to that of Ref.~\cite{ngs}. Moreover the event rate of 
Ref.~\cite{ball} above $\sim 20 \ GeV$ is about 25 \% of that of 
Ref.~\cite{ngs} (see Fig.~\ref{fig:spettri}). Therefore the background 
of $\pi^o$ due to NC will be reduced by the same amount, looking for 
$\tau \rightarrow e \nu \nu$ decay.

In the following, the beam from Ref.~\cite{ball} and 5 years of data
taking are assumed.

\section{$\tau$ appearance searches}
Tau appearance search is performed on the basis of kinematical
identification of the $\tau$ decay. The $\tau \rightarrow e \nu \nu$ is 
the channel favourite for this search due to the low background level 
and the the good electron identification capabilities of the TRD. It is 
worth noting that in the region of atmospheric anomaly the oscillation 
probability is $50 \div 100$ times higher than expected in NOMAD. As a 
consequence a much lower background rejection power is required.

In order to check the overall \NOE performances, a complete chain of
event simulation and analysis have been performed. Event generators
including Fermi motion, $\tau$ polarization and nuclear rescattering 
inside the nucleus have been used to simulate deep inelastic, quasi 
elastic and resonance interactions.

Generated events are processed by a GEANT based MonteCarlo in which 
calorimeter and TRD geometrical set-up are described in detail,
down to a scale of a few $mm$. Fiber attenuation length, Birks 
saturation, photoelectron fluctuations and readout electronics non 
linearities for both TRD and calorimeter have been taken into account.
DST of processed events ($\tau \rightarrow e \nu \nu$, $\nu_\mu$ NC
and $\nu_e$ CC) have been produced and analysed.

Electron identification is performed by looking for high energy releases 
in the TRD and in the calorimeter elements in fully contained events. 
The electron candidate is the one that maximizes collected energy 
in a $5^\circ$ cone centered at the interaction vertex. Electron direction 
is reconstructed by weighting hit position with collected energy. With 
present algorithms an angular resolution of $0.6^\circ$ and a 
$180 \ MeV/c$ resolution on the measurement of transverse momentum are 
achieved.

The remaining part of the event is used to reconstruct the hadronic
component. The obtained resolution on the measurement of transverse
momentum is $420 \ MeV/c$.
 
\begin{figure}
\hspace{+3.5cm}
\epsfig{file=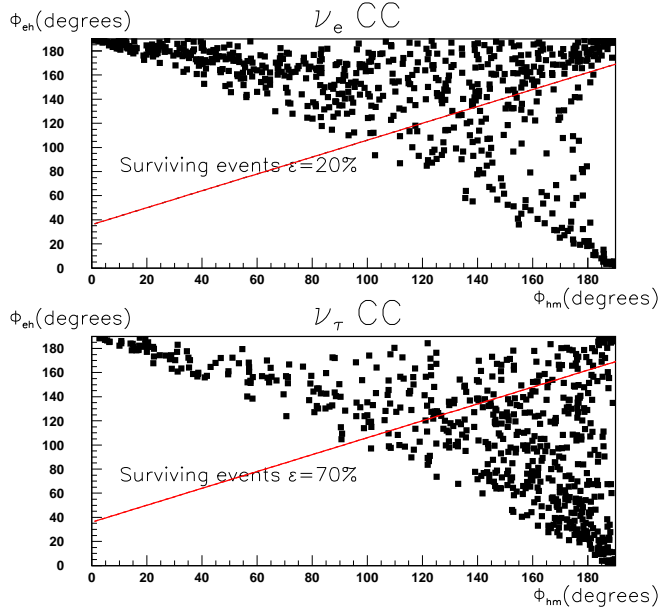,width=8.5cm,bbllx=31bp,bblly=150bp,bburx=550bp,bbury=630bp}
\vspace{-0.5cm}
\caption{\label{fig:fifi} Effects of $\phi_{e-h}$ $\phi_{m-h}$ cut 
on background ($\nu_e$) and signal ($\nu_\tau$).}
\end{figure}

Topological cuts on the electromagnetic shower are applied to reject 
$\nu_\mu$ NC events with $\pi^o$ faking electrons. Work is in progress 
to improve the reconstruction efficiency and $\nu_\mu$ NC rejection.
Additional cuts are performed to reduce the background:

\begin{itemize}
\item the total reconstructed energy $E_{tot} < 15 \ GeV$,
\item the electron energy $E_{e.m.} \ge 1.5 \ GeV$,
\item the component of electron momentum perpendicular to the hadronic
      jet direction $Q_{lep} \ge 0.75 \ GeV/c$,
\item the transverse mass 
      $M_T = \sqrt{4 p_T^e p_T^m \sin^2(\phi_{e-m}/2)} < 2 \ GeV$,
\item $\phi_{e-h}$ $\phi_{m-h}$ correlation as shown in 
      Fig.~\ref{fig:fifi}.
\end{itemize}

Analysis cut efficiencies and residual background percentages are 
reported in Table~\ref{tab:effi}. In Table~\ref{tab:tab2} the 
expected signal is reported for different values of $\Delta m^2$.

\section{Neural network for the ratio NC/CC}
\begin{figure}
\hspace{3.2cm}
\epsfig{figure=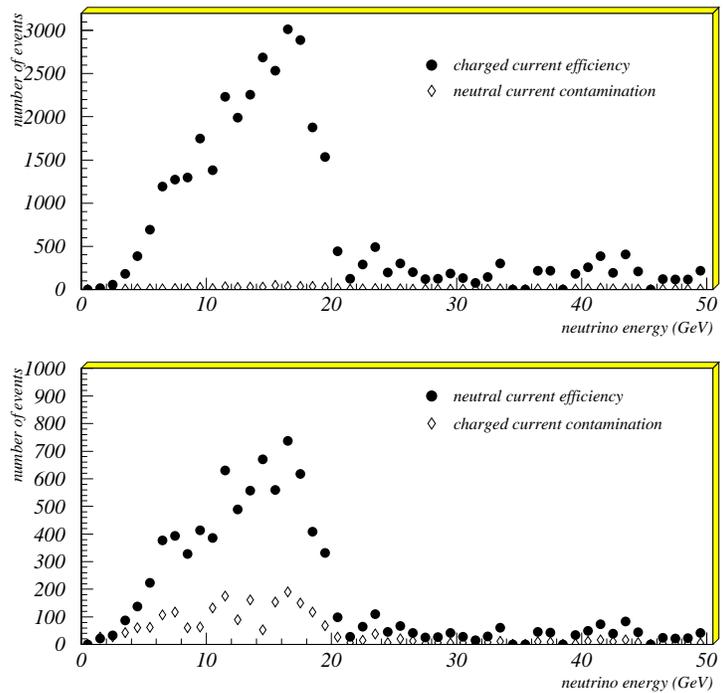,width=10.5cm}
\caption{CC and NC signal. The contamination is also shown. 
  \label{fig:orlo}} 
\end{figure}

The procedure to perform this measurement is well known \cite{noe6}. 
The identification of charged and neutral current events has been 
improved by means of a neural network. The algorithm uses 24 
topological, geometrical and calorimetric event parameters as input. 
The network has been trained with $\nu_\mu$ CC and NC MonteCarlo 
events with a neutrino energy uniformly distributed in the range 
$0 \div 50 \ GeV$. The results of this analysis are shown in 
Fig.~\ref{fig:orlo}.

The $\mu$ event recognition is mainly dependent on the track length 
and the NC signal is still affected by a significant CC contamination.
We hope to improve the recognition efficiency by tuning the neural 
network and the input parameters.

\begin{table}[b]
 \begin{tabular}{|l|c|c|c|} \hline
 Cut & Res. NC & Res. beam $\nu_e$ & eff. $\tau \rightarrow e\nu\nu$ \\\hline
Event containement          & 79 \% & 67 \% & 79 \% \\\hline
Track in two views with large & & & \\
released energy
                            & 19 \% & 88 \% & 91 \% \\\hline
Compact shower starting from the vertex, & & & \\
topology ($\pi^o$ without initial conversion) 
                            & 11 \% & 88 \% & 89 \% \\\hline
 $E_{tot} < 15 \ GeV$       & 94 \% & 45 \% & 97 \% \\\hline
 $E_{e.m.} \ge 1.5 \ GeV$   & 55 \% & 90 \% & 84 \% \\\hline
 $Q_{lep} \ge 0.75 \ GeV/c$ & 25 \% & 75 \% & 70 \% \\\hline
 $M_T < 2 \ GeV$            & 57 \% & 89 \% & 87 \% \\\hline
 $\phi_{e-h} \ \phi_{m-h}$ correlation 
                            & 35 \% & 20 \% & 70 \% \\\hline
 \end{tabular}
\caption{\label{tab:effi}Details on analysis cuts, residual 
background and signal recognition efficiency.}
\end{table}

\begin{table}[t]
 \begin{center}
  \begin{tabular}{|c|c|c|} \hline
 $\Delta m^2 \ (eV^2)$ & Background events & Signal 
($\tau \rightarrow e\nu\nu$ events) \\\hline
 $5 \times 10^{-3}$    & $2.7 \ \nu_e CC + 1.6 \ NC = 4.3$ &19 \\
 $3 \times 10^{-3}$    & $2.7 \ \nu_e CC + 1.9 \ NC = 4.6$ & 7 \\
 no oscillation        & $2.7 \ \nu_e CC + 2.0 \ NC = 4.7$ & 0 \\\hline
 \end{tabular}
 \caption{Estimated background and expected signal in 5 years 
 of data taking ($1.5 \times 10^{20}$ p.o.t. and $14000 \ \nu_\mu$ 
 CC in TRD target). \label{tab:tab2}}
 \end{center}
\end{table} 

\section{Conclusions}
The combined use of two subdetectors (TRD and CAL) allows to search for
$\tau$ appearance signal for events generated in the $2.5 \ kton$ TRD 
target where electron identification, vertex and kinematics 
reconstruction are performed at best. Nevertheless the whole $8 \ kton$
mass can be exploited for disappearance oscillation tests.

\begin{figure}
\hspace{2.8cm}
\epsfig{figure=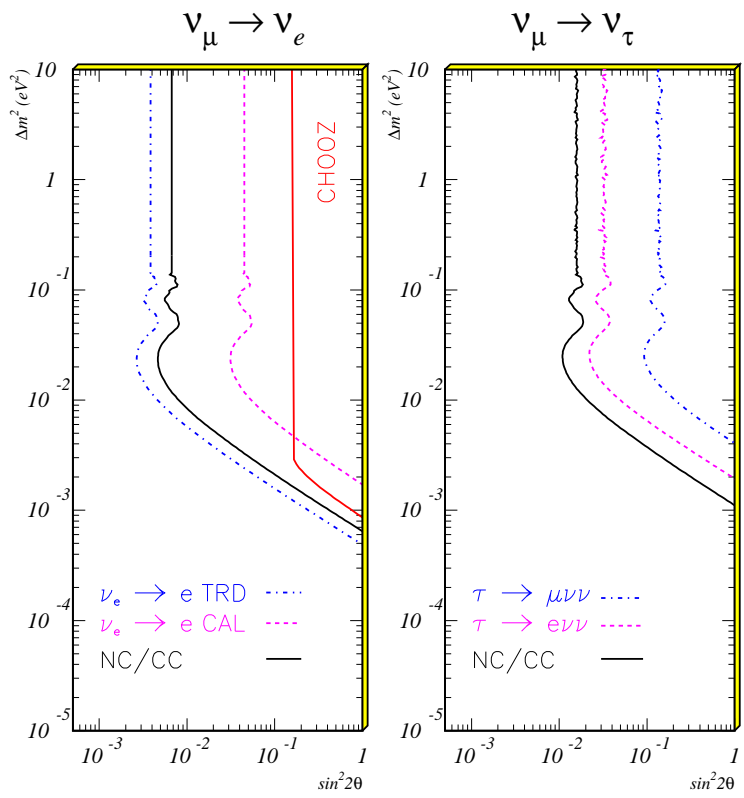,width=11cm}
\caption{\NOE sensitivity to $\nu$ oscillations. \label{fig:finale}} 
\end{figure}

Such measurements can be carried out at the same time by using an
appropriate neutrino beam. The feasibility of both measurements
has been demonstrated by means of full analyses. In 
Fig.~\ref{fig:finale} the sensitivity of \NOE experiment to $\nu$ 
oscillations is shown.

\end{document}